\documentclass[cpp,a4paper,fleqn%
]{w-art}
\usepackage{times,cite,w-thm}
\usepackage[utf8x]{inputenc}
\usepackage[english]{babel}
\usepackage[]{graphicx}
\usepackage{amsmath}
\usepackage{amssymb}
\usepackage{color}

\usepackage{ifpdf}

\ifpdf
\usepackage{epstopdf}
\fi
\begin{document}
\DOIsuffix{theDOIsuffix}
\Volume{XX}
\Month{ZZ}
\Year{YYYY}
\pagespan{1}{}
\Receiveddate{XXXX}
\Reviseddate{XXXX}
\Accepteddate{XXXX}
\Dateposted{XXXX}
\keywords{electron-hole bilayers, exciton formation, Wigner crystallization, complex plasma}



\title[Phase diagram of bilayer electron-hole plasmas]{Phase diagram of bilayer electron-hole plasmas}


\author[J. Schleede]{Jens Schleede\inst{1,}%
  \footnote{E-mail:~\textsf{schleede@physik.uni-greifswald.de}}
 }
\author[A. Filinov]{Alexey Filinov\inst{2,}\inst{3,}
\footnote{E-mail:~\textsf{filinov@theo-physik.uni-kiel.de}}
 }
\author[M. Bonitz]{Michael Bonitz\inst{2}
}
\author[H. Fehske]{Holger Fehske\inst{1}
}
\address[\inst{1}]{%
  Institute for Physics,
  Ernst-Moritz-Arndt Universit\"at,
  Greifswald, Germany}
\address[\inst{2}]{%
  Institute for Theoretical Physics and Astrophysics,
  Christian-Albrechts-Universit\"at,
  Kiel, Germany}
\address[\inst{3}]{%
  Joint Institute for High Temperatures RAS, Izhorskaya Str.~13, 125412 Moscow, Russia}


\newcommand{\vect}[1]{{\mathbf{#1}}}

\begin{abstract}
We investigate exciton bound-state formation and crystallization effects in two-dimensional electron-hole bilayers.
Performing unbiased path integral Monte Carlo simulations all quantum and Coulomb correlation effects are treated on first principles.
We analyze diverse pair distribution functions in dependence on the layer separation, particle density and hole-to-electron mass ratio and derive a schematic phase diagram for the neutral mass-asymmetric bilayer system.
Our simulations reveal a great variety of possible phases namely an exciton gas, an exciton crystal, an electron-hole liquid and a hole crystal embedded in an electron gas.
\end{abstract}

\maketitle                   


\section{Introduction}\label{intro}

Strongly correlated classical and quantum two-component systems are of high interest in many fields, for a recent overview see \cite{sccs_2011}.	
Examples include electron-hole plasmas in semiconductors, e.g. \cite{haug-koch-buch,exc,d1,d2}, dense electron-ion plasmas (warm dense matter) and complex or dusty plasmas \cite{meichsner_cpp12,melzer_cpp12, bonitz_rpp}. While the simple one-component plasma (or jellium) model of a Coulomb system has been studied in detail it disregards many fundamental properties, in particular the formation and phases of bound states. Much less is known about ``real'' two-component plasmas.
 Although a number of studies have been performed in recent years, e.g. \cite{de_palo_excitonic_2002,kalia,senatore_correlation_2003,vsfilinovehplasma}, many questions remain. In particular, the phase diagram and its dependence on the mass ratio of the two components is known only qualitatively. 

Here, we analyze these questions by considering the simplest realization of a neutral two-component plasma: A bilayer system with spatially separated light negative and heavy positive charges,  in the following referred to as electrons (``$\text{e}$'') and  holes (``$\text{h}$''). 
%
At low temperature and large mass anisotropy, $M=m_h/m_e \gg 1$, the (classical) heavy component has a tendency to form a Wigner crystal whereas the light component forms a neutralizing background.
Thus, a strong (weak) coupling situation is realized for the heavy (light) component, which can be quantified in terms of the one-component coupling parameter, $\overline{\Gamma}_{\alpha}=[e^2 /\epsilon\bar{a}]/[\hbar^2/m_\alpha \bar{a}^2]=\bar{a}/a_B$, where $m_\alpha$ are the effective particle masses $(\alpha=\text{e},\text{h})$, $e$ is the elementary charge, $\epsilon$ the dielectric constant, and $\hbar$ denotes the reduced Planck constant.
For each species, $\overline{\Gamma}_{\alpha}$ depends on $m_\alpha$ (via the Bohr radius $a_B=\hbar^2\epsilon/ m_\alpha e^2$) and the particle number density $\rho \sim 1/\bar{a}^\text{D}$ ($\bar{a}$ is the average inter-particle distance and $\text{D}$ the dimensionality of the system).
When $\text{e}$-$\text{h}$ bound state (exciton) formation takes place, the residual (screened) interaction between these almost neutral complexes weakens, i.e. $\overline{\Gamma}$ is reduced, up to the point where a crystallization into a Wigner solid is prevented.
Generally the formation of few-particle bound states is limited to low temperatures. Their correct theoretical description calls for a full quantum-mechanical treatment, in particular if temperature and density effects, as well as their stability against quantum fluctuations is addressed for low-dimensional systems.
This provides the motivation of the present work: After giving a brief overview on existing results on e-h bilayers,
we perform a first-principle numerical analysis of thermodynamically stable phases of a two-component Coulomb plasma in the bilayer geometry with spatially separated electrons and holes, and map out major regions of the phase diagram.
Thereby the focus is on the inter-layer and intra-layer particle correlations, leading to bound state formation and charge ordering, respectively.  

\section{Electron-hole bilayers}\label{overview}
The main advantage of a bilayer system is the easy control of the coupling between the layers via the layer separation $d$. 
Actual realizations range from electron-ion quantum plasmas~\cite{sccs_2011,vsfilinovehplasma} to electron-hole systems in semiconductors~\cite{haug-koch-buch,vsfilinovsemi}.
In the latter spatially separated electrons and holes can be realized in double quantum well heterostructures~\cite{exc,d1,d2}.
Recent advances in fabrication technology (lithography techniques and molecular beam epitaxy) allow for a good control of the electron/hole density and the spatial inter-well distance.
Both classical and quantum electron-hole bilayers have been a subject of active research during the last decades~\cite{ludwig2007,peter1,peter2,de_palo_excitonic_2002,senatore_correlation_2003}.
{\em Classical bilayers} (corresponding to high temperature and low density) have been investigated by molecular dynamics simulations to reveal their static and dynamical properties~\cite{peter1,peter2}.
The phase diagram has been explored as a function of the layer separation $\tilde{d}\equiv d/\bar{a}$, with the principle result that for $1 \leq \tilde{d} \leq 2.5$ the inter-layer correlations stabilize the Wigner crystals in both layers (the melting temperature is increased) compared to a single layer Wigner crystal.
For {\em quantum bilayers}, at zero temperature, the static and dynamic dielectric response has been analyzed in Refs.~\cite{q1,q2,q3}.
The presence of a second (oppositely charged) layer was found to enlarge the region where a crystalline phase exists in the  $d-r_s$ plane to higher densities, $\rho_c(d) \geq \rho_c(d\rightarrow \infty)$, or $r_s(d) \leq r_s(\infty)$
($r_s\equiv \overline{\Gamma}_{e}$ denotes the Brueckner parameter).
The critical value of the parameter $r_s$, characterizing the critical particle density, changes from $r_{s,c}\approx 37$ for independent layers~\cite{tan} to 
$r_{s,c} \approx 20$ for a bilayer.
Quantum melting of a Coulomb crystal is caused by the zero-point motion of the particles around their lattice sites.
Similar to the classical case, these fluctuations are suppressed by attraction 
to the particles residing in the second layer, in particluar if they also form a  lattice.
In this way, a stack of two Wigner lattices appears to be more stable then a single layer crystal.

The second component plays also a prominant role when electron-hole bound states are formed, e.g.\ for systems of spatially indirect excitons.
Then the quantum bilayer is similar to a 2D system of dipoles.
The transition from Coulomb (dominant at $\tilde{d} \sim 1$) to dipolar ($\tilde{d} < 1$) correlations shifts the critical density of the solid-gas phase transition to a lower value~\cite{dipoles}, which scales with the dipole moment (or the layer separation) as $\rho_c a_B^2=1/\pi r_s^2=\tilde{d}^4/290$.
In the dipole approximation ($\tilde{d} \ll 1$), the bilayer phase diagram should be similar to that for bosonic dipoles in 2D and includes superfluid and normal gases~\cite{bkt} as well as a dipolar solid, if the composite particles obey Bose statistics.
A more accurate treatment that accounts for the internal structure of bound states and allows to release the restriction $\tilde{d} \ll 1$, was recently performed, see Ref.~\cite{jens}.
There it was shown that the a realistic finite layer width restores the $1/r$-Coulomb scaling at small distances which makes an exciton solid unstable at high densities similar to the case of a 2D Wigner crystal.
Besides macroscopic systems, {\em mesoscopic bilayers} with full account of Fermi statistics have been studied recently~\cite{karsten}.
Such systems are of particular relevance for experiments with carriers confined by an external trap potential, e.g. electron-hole systems in coupled quantum dots, where both the distance between the dots and the in-plane carrier density are tunable parameters.
In contrast to bulk systems, in a mesoscopic bilayer, a change of $d$ leads to a sequence of structural transitions and shell structure formation~\cite{karsten} similar to complex plasmas \cite{melzer_cpp12}.

Another important aspect, which has not been studied in detail so far, concerns the influence of the mass-asymmetry $M$ between the bilayer charge carriers. 
A mass ratio $M \neq 1$ is expected to have a significant effect on the phase diagram, as the zero-point energies of light electrons and heavy holes differ, which in turn will influence inter-particle correlations~\cite{moud,ludwig2007}.
Depending on the particle density, the light component (electrons) can be in a homogeneous Fermi gas phase, while the heavier component (holes) can form a Wigner lattice when the mass ratio exceeds a critical value, $M\geq M_c$ (with $M_c\gtrsim 4$ for typical electron densities in semiconductors which is about a factor $20$ lower than in a $3D$ plasma \cite{vsfilinovehplasma}).
The critical value $M_c$ can be tuned by the layer separation $d$~\cite{ludwig2007}.
This is in favor of Wigner crystallization (at least for a heavy component) in a two-component Coulomb system.
In contrast, in a 2D one-component system (electron gas) crystalliztation  (which is expected to appear for $r_s \gtrsim 37$~\cite{tan}) could not yet be observed~\footnote{A recent experimental demonstration of Wigner islands of electrons in a mesoscopic structure covered with a helium film deals with the near-classical limit~\cite{rous}.} since the typical particle densities are too high ($r_s \leq 10$), and quantum fluctuations destroy any spatial ordering.

In the following, we present a more systematic analysis of {\em macroscopic bilayers}, by considering a wider range of densities/inter-layer separations and mass ratios. We identify different structural phases by analyzing the pair distribution functions 
and reconstruct the phase diagram. For this purpose we adopt the path integral Monte Carlo (PIMC) technique for simulations of spatially separated electron-hole systems~\cite{ludwig2007}.
We note that quantum effects between electrons and holes are fully taken into account by the PIMC formalism based on the stochastic sampling of the $N$-particle density matrix.

\section{Theoretical model}
A wide class of electron-hole bilayers can be described by the Hamiltonian
\begin{align}
  H &= H_\text{e} + H_\text{h} - \sum_{i=1}^{N_\text{e}}\sum_{j=1}^{N_\text{h}} \frac{e^2}{\epsilon \sqrt{|\vect{r}_{i,\text{e}}-\vect{r}_{j,\text{h}}|^2 + d^2}}\;,\label{eq1}\\
  H_{\alpha} &= - \sum_{i=1}^{N_{\alpha}} \frac{\hbar^2\nabla^2_{\vect{r}_{i,\alpha}}}{2 m_{\alpha}}
         + \sum_{i<j}^{N_{\alpha}} \frac{e^2}{\epsilon \left|\vect{r}_{i,\alpha}-\vect{r}_{j,\alpha}\right|}\;.\label{eq2}
\end{align}
The last term in Eq.~(\ref{eq1}) mimics the coupling between the electrons and holes residing in different layers. 
Obviously, the ground-state and thermodynamic properties of the system depend on the following experimentally accessible and managable parameters: 
(i) the hole-to-electron mass ratio $M=m_\text{h}/m_\text{e}$, 
(ii) the inter-layer separation $d\equiv d/a_B$,
(iii) the in-plane particle density $\rho a_B^2=1/\pi r_s^2$, and 
(iv) the temperature $k_B T$ (measured in Hartree: $\text{Ha} = \hbar^2 / m_\text{e} a_B$).
In the simulations we considered a bilayer system with $N_\text{e} = N_\text{h} = 64$ electrons and holes.
The particle species are confined to their respective two-dimensional layer of zero width, separated by $d$.
To simulate an infinite system we use a rectangular simulation cell with periodic boundary conditions.
The dimensions of the cell with $L_y = \sqrt{3}{L_x}/2$ and the number of particles were chosen to fit the symmetry of a hexagonal lattice, once a Wigner crystal is formed.
The temperature was kept fixed to $k_B T = 1/3000\text{ Ha}$.
The other parameters ($M, r_s, d$) were varied, each in the range 
$1\ldots 100$.
More details regarding the setup of the PIMC simulations can be found in Ref.~\cite{ludwig2007}.
In order to identify different structural phases we exploit the pair distribution function between different species ($\alpha\beta=\text{ee},\text{hh},\text{eh}$)
\begin{align}
  g_{\alpha\beta}(r) = C_{\alpha\beta} \sum_{i=1}^{N_\alpha} \sum_{j=j_{\alpha\beta}+1}^{N_\beta}%
              \langle \delta(\left|\vect{r}_{i,\alpha}-\vect{r}_{j,\beta}\right|-r) \rangle \, ,
              \label{gr} \\
 C_{\text{ee}(\text{hh})}=\frac{2}{N_{\text{e}(\text{h})}(N_{\text{e}(\text{h})}-1)}, \; j_{\text{ee}(\text{hh})}=i, \quad C_{\text{eh}}=\frac{1}{N_\text{e} N_\text{h}}, \; j_{\text{eh}}=0 \, .
\end{align}
In particular, the formation of a Wigner solid in the electron (hole) layer is ascribed to a jump in the ratio of the first maximum and the next minimum in $g_{\text{ee}(\text{hh})}$, i.e. 
\begin{align}
    \Delta=\max\limits_{r< 2r_c}[g_{\alpha\alpha}(r)]\,\, / \min\limits_{r< 2r_c}[g_{\alpha\alpha}(r)] \, ,
    \label{delta}
\end{align}
where $r_c=\max[g_{\alpha\alpha}(r)]$ is the radius of the first correlation shell.
This criterion has been found sensitive to a solid-liquid/gas phase transition~\cite{melt1,melt2}.
Furthermore, to identify phases composed of excitons we use the fraction of bound electron-hole states, which can be estimated by the inter-layer pair distribution function $g_\text{eh}$ 
(for more details see \cite{bsf,bjerrum}),
\begin{align}
    X = \frac{N^b_\text{eh}}{N^b_\text{eh}+N^c_\text{eh}} 
    \approx
    \frac{\displaystyle \int_0^{r_b} r[g_\text{eh} - 1] \,\text{d}r}
         {\displaystyle \int_0^{r_b} r g_\text{eh} \,\text{d}r} \, .
    \label{bsf}
\end{align}
Here $r_b$ is the second zero of $r [g_\text{eh}(r)-1]$ on the right-hand side of the exciton peak located near $r=0$.

\section{Results and discussion}
\subsection{Characterization of different bilayer phases}
\begin{figure}[hp]
  \newcommand{\mywidthfig}{0.96\linewidth}
  \newcommand{\mywidthlabel}{0.03\linewidth}
  \newcommand{\myfigure}[2]{
    \begin{minipage}[t]{\mywidthfig}\vspace{0pt}
      \includegraphics[width=\textwidth]{#1}
    \end{minipage}
    \begin{minipage}[t]{\mywidthlabel}\vspace{0pt}
       \hfill {#2}
    \end{minipage}
  }
  \myfigure{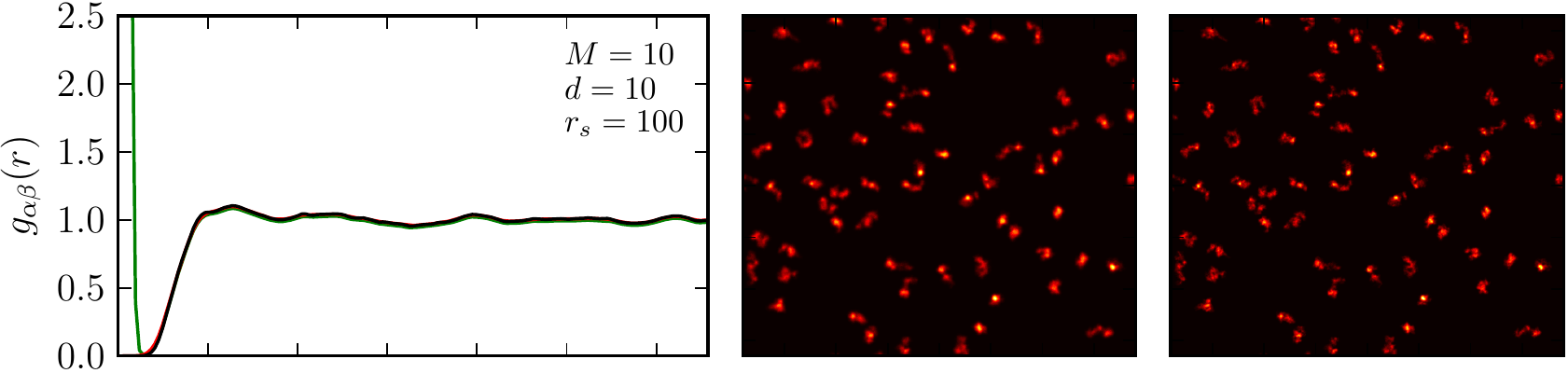}{a)}
  \myfigure{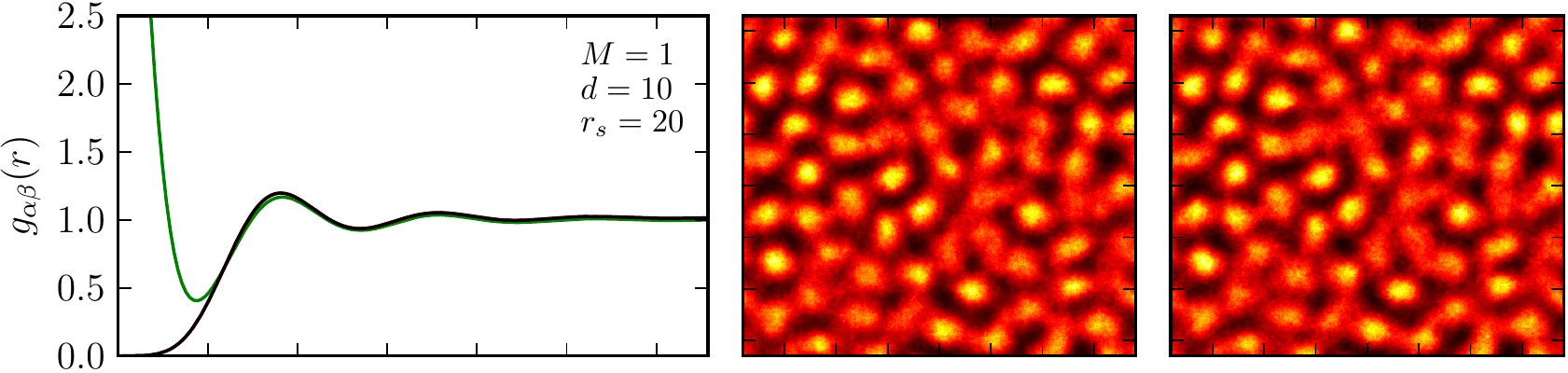}{b)}
  \myfigure{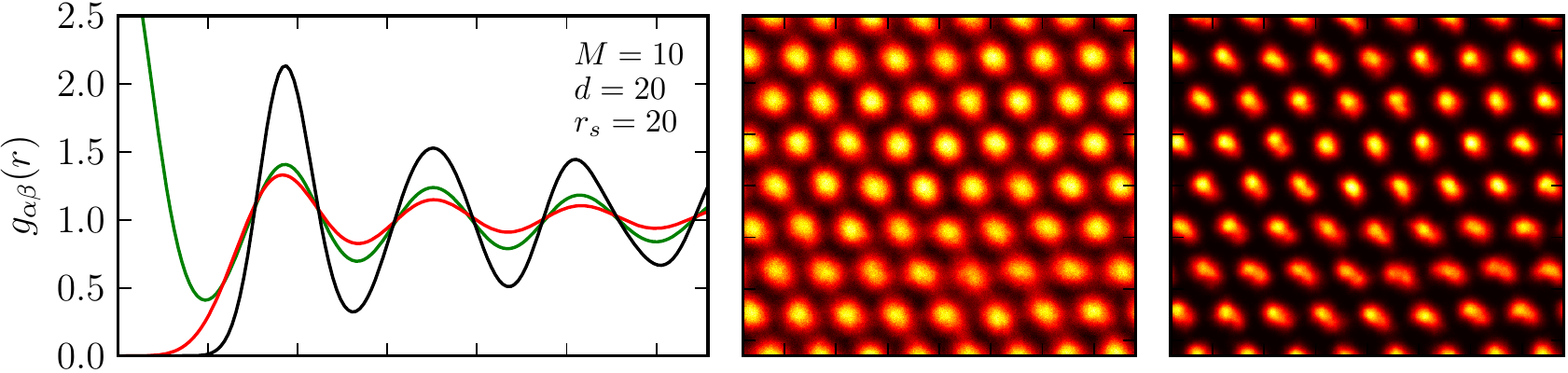}{c)}
  \myfigure{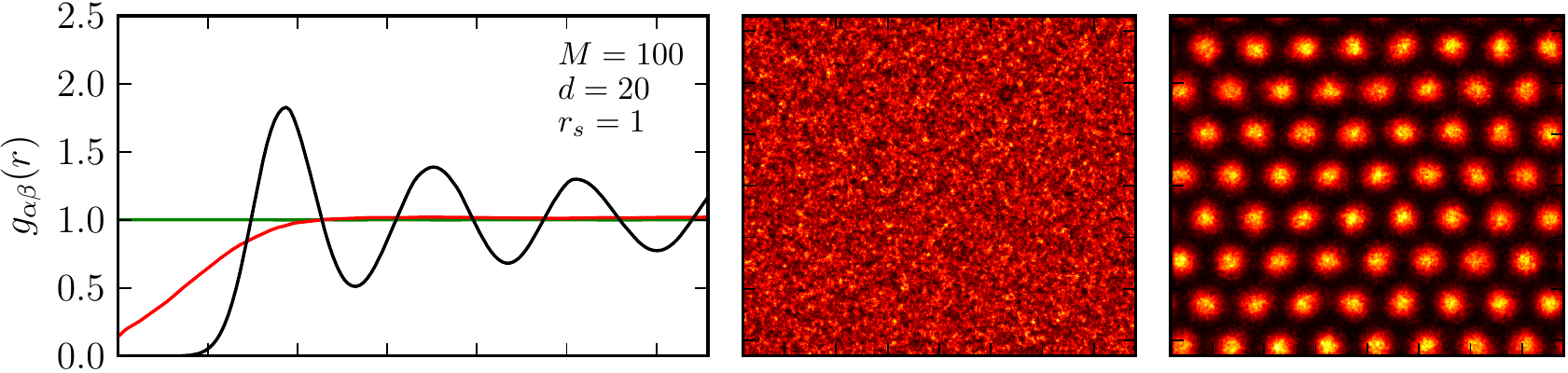}{d)}
  \myfigure{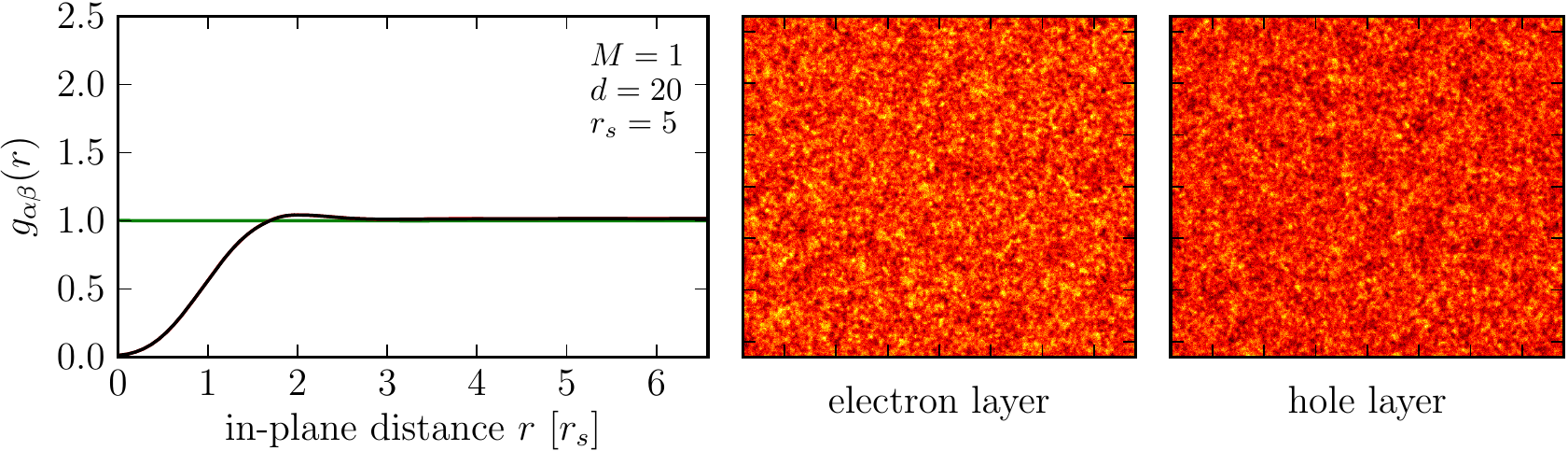}{e)}
  \caption{
    {\em Left column}: Hole-hole ($g_{\text{hh}}$, black lines), electron-electron ($g_{\text{ee}}$, red lines) and electron-hole ($g_{\text{eh}}$, green lines) pair distribution functions (averaged over $100000$ Monte Carlo steps).
    {\em Middle/right columns}: Snapshots of the instantaneous particle density (averaged over $10000$ MC steps) in the electron and hole layer, respectively.
    Varied parameters are mass ratio $M$, inter-layer distance $d$, and the Brueckner parameter $r_s$.
    The three upper rows (from top to bottom) typify the excitonic phase: a) a classical exciton gas, b) a degenerate exciton gas and c) an exciton crystal.
    The last two rows (no e-h bound states) mark d) a hole Wigner crystal and e) an electron-hole plasma.
    All PIMC simulations were performed with $N_\text{e} = N_\text{h} =64$ particles in each layer.
    The temperature was kept fixed at $k_{B}T=1/3000 \textrm{ Ha}$.
  }
  \label{fig:snapshots}
\end{figure}

Fig.~\ref{fig:snapshots} presents our simulation data for the pair distribution functions (left panels) and corresponding density ``snapshots'' (right panels) for characteristic configurations of the electron-hole bilayer system.

{\em Classical exciton gas.}
Fig.~\ref{fig:snapshots}a exemplifies the excitonic gas phase as realized for a mass-asymmetric bilayer with $M=10$ and an inter-layer distance much smaller compared to the intra-layer particle distance ($d/r_s=0.1\ll 1$).
Therefore we are in a regime where the dipole approximation is valid.
Indeed, $g_{\text{eh}}(r)$ is sharply peaked at $r=0$ reflecting the formation of bound states (indirect excitons).
A comparison of the snapshots of the (instantaneous) particle density in each layer shows that all particles are bound pairwise (cf. the uppermost two right-hand panels).
The structureless form of $g_{\text{ee}(\text{hh})}$ shows that the excitons interact only weakly, i.e., they move almost freely and form a gaseous phase.
Expressed in terms of the dipole coupling parameter $D$ (see Ref.~\cite{bkt}), this regime is characterized by $D=d^2(M+1)/\sqrt{\pi} r_s\sim 6.2$, and we expect a dipolar gas with moderate correlations and a superfluid transition temperature $T_c[E_0] \equiv k_B T_c/E_0\sim 1.25$ 
(cf. the $D$-dependence of $T_c$ in Ref.~\cite{bkt} scaled in units $E_0=\hbar^2 \rho/(M+1)$).
Since $T_c$ is by factor $10^{-2}$ smaller than the temperature simulated in Fig.~\ref{fig:snapshots}a, where $T[E_0]\equiv k_B T/E_0=(k_B T/{\rm Ha}) (M+1) \pi r_s^2\sim 115$, the exciton gas is essentailly {\em classical}, and effects of Bose/Fermi statistics are irrelevant.
The situation does not change much if $M=1$, because the degeneracy of the exciton gas remains low.

{\em Degenerate exciton gas/liquid.}
We now consider the mass-symmetric case and increase the particle density (reduce $r_s$), see Fig.~\ref{fig:snapshots}b. 
Since for $d=10$ and $r_s=20$ the condition $d/r_s<1$ still holds, the excitons still behave as dipoles.
This is supported by an analysis of the microscopic effective exciton-exciton interaction potential~\cite{jens}.
For the parameters used in Fig.~\ref{fig:snapshots}b, we have $D\sim 5.6$ and $T_c[E_0]\sim 1.3$.
The simulated temperature, $T[E_0]\sim 0.84$, is now below $T_c$.
Accordingly the exciton gas is superfluid.  In the quasi-2D bilayer excitonic system the superfluid transition will be of the Berezinskii-Kosterlitz-Thouless type~\cite{berez,kost} and is associated with the emergence of topological order, i.e. the pairing of vortices with opposite circulation. 
The correlations of the order parameter (for example, the one-particle density matrix) decay algebraically in space.  A true long-range order is recovered only at $T=0$. This scenario is applicable to a wide variety of quasi two-dimensional systems. Recent experimental realizations of degenerate dipolar atomic gases~\cite{cr}, bosonic molecules~\cite{mol}, and indirect excitons in quantum wells~\cite{exc,d1,d2} have stimulated theoretical analyses of their quantum many-body properties. The present bilayer system provides an interesting test case.
In our system, the onset of degeneracy and spatial coherence in both layers (which leads to the superfluid response) is reflected in the behavior of the pair distribution functions. The incipient overlap of the excitonic wavefunctions (around $r\sim r_s$) can be noted from the particle density plots (right-hand panels of Fig.~\ref{fig:snapshots}b). Simultaneously, we observe the build-up of short-range correlations which is indicated by the few (weak) oscillations in $g_{\text{ee}(\text{hh})}$. The density snapshots point to a local hexagonal ordering: Though noticeable spatial correlations exist between nearest neighbor and even next-nearest neighbor excitons, true long-range order does not appear.
Such behavior is typical for correlated liquids with repulsive interactions. Finally, we note, that the correlations not only influence the static structural properties, but they significantly modify also the quasi-particle excitation spectrum (leading to the formation of a phonon-maxon-roton branch). The latter has a strong effect on the superfluid transition temperature (and stability of the superfluid phase to thermal fluctuations), in particular, in 2D dipolar systems~\cite{bkt}.

{\em Excitonic crystal.}
An interesting situation appears when the inter-layer and intra-layer correlations are of the same order.
If $d \sim r_s$, the bilayer system does not behave as a pure dipolar system.
For the parameters used in Fig.~\ref{fig:snapshots}c we get $D\sim 124$ and $T[E_0]=4.6$, i.e., we are in the classical domain of the phase diagram determined in Ref.~\cite{bkt}, and far away from the quantum crystallization transition observed at $D\sim 17$~\cite{dipoles}.
In this case, a crude estimate of the melting temperature of a dipolar crystal follows from the classical dipole coupling parameter $\Gamma_D\sim 60$~\cite{kalia}, yielding $k_B T/{\rm Ha}=d^2/(\Gamma_D r_s^3)\sim 8 \cdot 10^{-4}|_{d=r_s=20}$.
This means the simulated temperature ($k_B T=1/3000 \textrm{ Ha}$) is $2.5$ times below the melting temperature and, indeed, we observe an almost perfect hexagonal lattice in both layers. 
Obviously the bilayer geometry plays a crucial role, and we directly observe the stabilizing effect of the second layer mentioned in the introduction: 
A single electron layer would form a Wigner crystal only at $r_s\gtrsim 37$~\cite{tan}, whereas Wigner crystal formation in a bilayer sets in at significantly higher densities, e.g. at about $r_s\approx 20$ for the mass-symmetric case $M=1$, see Refs.~\cite{q1,q2,q3}.
In qualitative agreement, our data for a bilayer with heavy holes yield an exciton crystal already for $r_s \gtrsim 10$.
This state is characterized by strongly oscillating pair distribution functions 
in both layers.
The observed difference in the peak heights of $g_{\text{hh}}$ and $g_{\text{ee}}$ results from the mass-asymmetry.
Since the lighter electrons have a larger zero-point energy, they are not as localized as the holes around the hexagonal lattice sites.
Furthermore we can test the appearance of the present bilayer exciton solid against the predictions of the exciton approximation~\cite{jens} (here the analysis has been performed for a mass ratio $M=2.46$, which corresponds to the paramaters of ZnSe-based quantum wells).
The excitonic phase diagram derived in Ref.~\cite{jens} predicts that such kind of exciton solid can exist only in a finite density interval, which depends on the exciton dipole moment (and on the inter-layer separation, in case of a bilayer).
For $d=20 a_B$, the density simulated in Fig.~\ref{fig:snapshots}c  ($r_s=20$) fits in the exciton solid density window, $9.4 \leq r_s \leq 46$, estimated from Fig.~7 in~\cite{jens}.

{\em Hole crystal.}
We now turn to the case where electrons and holes are weakly correlated.
This happens at high in-plane particle densities (and large layer distances) when the Coulomb attraction between electrons and holes is screened to a large extent. 
Then the correlation energy within the layer, $E_c\sim 1/r_s$, overcomes the inter-layer binding energy, $E_B \sim 1/d$.
As a result the electron-hole pairs dissociate and the spatially separated electron and hole subsystems undergo structural phase transitions essentially independently. For the strongly mass-asymmetric case, zero-point fluctuations destroy any ordering in the layer containing the light electrons, while the heavy holes can form a Wigner lattice, see the results for $M=100$ in Fig.~\ref{fig:snapshots}d.
In the hole layer, the one-component quantum parameter scales with the particle density as $r_s^\text{h}=m_\text{h}/m_\text{e} r_s^\text{e}$, leading to $r_s^\text{h}|_{r_s^\text{e}=1}=100$ well above $r_{s,c}\approx 37$~\cite{tan}.
Formation of the hole crystal does not necessarily go along with a significant modulation of the electron density.
This can be seen from the density snapshot but also from the behavior of the e-h pair distribution function, which stays practically constant ($g_{\text{eh}} \approx 1$) for all distances.

{\em Electron-hole plasma}.
The hole crystal melts, of course, if we lower the mass of the holes.
For the mass-symmetric case shown in Fig.~\ref{fig:snapshots}e, a structureless electron (hole) plasma emerges in both layers.

\subsection{Phase diagram}
\begin{figure}[th]
  \includegraphics[width=\linewidth]{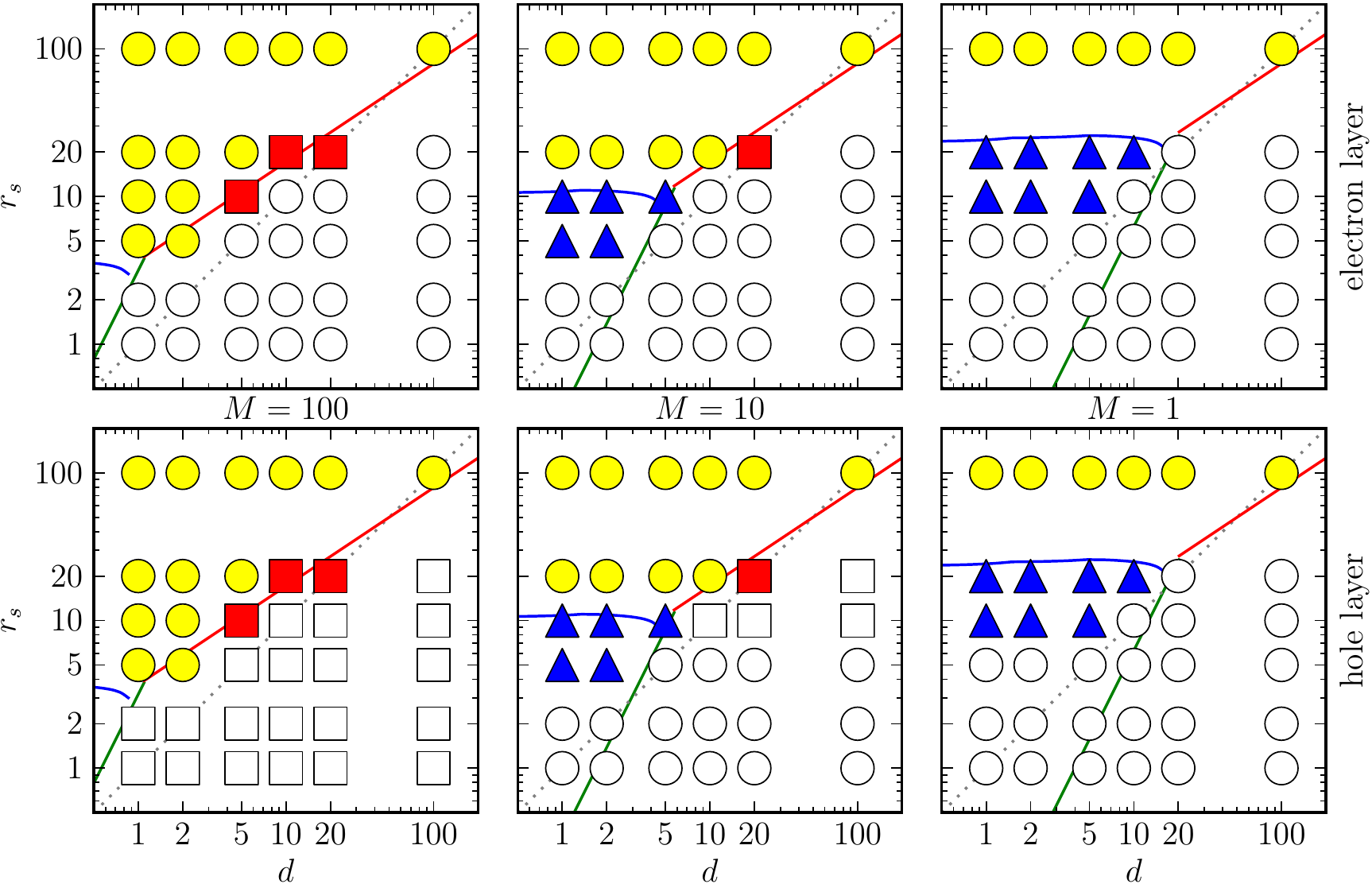}
  \caption{%
    Phase diagram of an electron-hole bilayer in the $d-r_s$ plane 
for mass ratios $M=100$ (left column), $M=10$ (middle), and $M=1$ (right).
    The upper (lower) row gives data for the electron (hole) layer. 
    Symbols denote different thermodynamic phases: crystal (squares), degenerate gas (triangles) and classical gas (circles).
    Bound states (indirect excitons) are marked by filled symbols.
    The superfluid phase for 2D dipole systems is located just below the blue curve (superfluid-normal phase boundary) and left of the green line (superfluid-dipole crystal boundary, corresponding to $D=18$) as taken from Ref.~\cite{bkt}.
    The degenerate exciton gas, shown as filled triangles, is in the superfluid state.
    The melting transition boundary of the classical dipole phase is mareked by the red line.
    The dotted line is a guide for the eye showing the $d-r_s$ diagonal.
    }
  \label{fig:phasediagram}
\end{figure}

After having  qualitatively characterized the various phases found in the electron-hole bilayer system, we now construct the phase diagram in the $d-r_s$ plane, see Fig.~\ref{fig:phasediagram}. 
The identification of the different states and configurations is based on the max-min ratio $\Delta$ and the bound state fraction $X$ given by Eq. (\ref{delta}) and (\ref{bsf}), respectively.
The Wigner crystal phase is found above a critical value $\Delta_c = 2.75$ and the excitonic phase was specified for $X > X_c = 0.33$.
Of course, these specified critical values yield only an estimate of the exact phase boundary in the thermodynamic limit.
The first general observation from the phase diagram is that, independent of the mass ratio $M$, excitonic bound states exist only for $r_s \gtrsim d$ and $r_s \gtrsim 5$.
These states are marked in Fig.~\ref{fig:phasediagram} by filled symbols. 
A similar observation was reported in Ref.~\cite{liu_exciton_1998}.

{\em Large mass-asymmetry.}
For $M=100$ and $r_s \geq d$ we observe some of the exciton phases discussed in the previous section (cf. the left most column in Fig.~\ref{fig:phasediagram}). 
In the neighborhood of $r_s=d$ a bilayer exciton crystal is formed (filled squares).
By lowering the density ($r_s=100$) the exciton crystal vanishes, as predicted by the dipole approximation valid in this regime. 
Once the inter-particle spacing $\bar{a}$ is increased, the dipolar potential energy $V\sim d^2/\bar{a}^3$ will decay faster than the zero-point kinetic energy $K\sim \hbar^2/m \bar{a}^2$.
Hence, at some point quantum fluctuations will destroy any spatial ordering and initiate a phase transition to a non-degenerate (classical) exciton gas (filled circles), far away from the diagonal $r_s=d$ (recall that large $r_s$ implies a dilute, weakly interacting system).
Below the diagonal ($r_s< d$), the excitons dissociate.
As a result the two layers are weakly coupled and exhibit essentially independent structural phases.
For the scanned range of $r_s$-values, $1\leq r_s < 100$, the Wigner crystal in the hole layer is the thermodynamically stable phase, due to the large hole mass. By contrast, the electron layer at high densities (small $r_s$) forms 
a degenerate gaseous phase up to $r_s = 5$.
Here an electron Wigner crystal is not observed, because the two points we have simulated ($r_s=20$ and $r_s=100$) are either below or well above the critical density $r_{s,c}\approx 37$~\cite{tan}.
For $r_s=100$, the simulated temperature $k_B T=1/3000 \textrm{ Ha}$ is not low enough to see Coulomb crystallization of a classical electron gas.
The corresponding classical Coulomb coupling parameter $\Gamma=(e^2/\bar{a})/k_B T=(1/r_s)/(k_B T/\textrm{Ha})=30$ is well below the critical value $\Gamma_c\approx 137$, required for crystallization of a classical electron gas.

{\em Moderate mass-asymmetry.}
If we decrease the mass ratio to $M=10$  the stability of the excitonic phase is not much affected (cf. the middle panels of Fig.~\ref{fig:phasediagram}). 
For $r_s>d$ and $r_s >5$ their phase-space domain remains almost the same. 
At $r_s=d=20$ a bilayer exciton crystal forms (see also Fig.~\ref{fig:snapshots}c).
Here we note the stabilizing effect of the lattice in the hole layer.
Due to the coupling of the electrons to the holes arranged in a regular pattern, the electron Wigner crystal is observed significantly below the corresponding critical value $r_{s,c}\approx 37$ (above the density) of a single layer.
Obviously, the hole lattice produces the same pinning effect as impurities studied in Ref.~\cite{tan}. In the latter case, in-plane impurities enhance the electron Wigner crystallization, i.e., they shift the transition density to $r_s\sim 7.5$.
The same scenario applies to bilayers with large inter-layer separation, provided the screening effect produced by the electrons on the hole-hole interaction can be neglected.
Our simulation show that at higher electron densities ($r_s\leq 10$) zero-point fluctuations may overcome the pinning effect of the hole lattice and the electron Wigner crystal melts on the diagonal $d=r_s$ (see middle column of Fig.~\ref{fig:phasediagram}, $d=r_s=10$).
At an inter-layer separation $d=5$ the melting of the hole lattice is accompanied by the formation of excitonic bound states. 
That is to say we observe a destabilization of the hole Wigner lattice by the electron layer due to screening effects.

{\em Mass-symmetric bilayer.} 
In the case $M=1$, obviously, all phase transitions are identical in both layers.
For $r_s<10$ and $r_s<d$ we find an electron-hole plasma, otherwise a (non-degenerate or degenerate) excitonic gas phase, see right panels of Fig.~\ref{fig:phasediagram}.
Note, that for the layer separation $d=1$, exciton bound states are stable for $r_s \geq 10$, in agreement with zero-temperature diffusion Monte Carlo results~\cite{de_palo_excitonic_2002}.
We can not confirm, however, the two-component plasma phase boundary reported in 
Ref.~\cite{de_palo_excitonic_2002}, simply because the corresponding region 
$r_s,d<1$ was not accessible by our PIMC simulations so far.
As noted before, the simulated temperature is not sufficiently low 
to observe a single-layer Wigner crystal at $d \gg 1$ and $r_s\geq 37$. 
A similar argument applies to the exciton crystal, which should 
exist for layer separations above a critical value $d \geq d_c$, where 
$d_c\approx 6 a_B$ for $M=2.46$~\cite{jens}.
In the vicinity of the excitonic Mott transition more refined simulations, including an accurate analysis of the fraction of ionized and bound states and a full account of the Fermi statistics \cite{schoof11} of electrons and holes  would be required.

\section{Summary and concluding remarks}
In this report, we analyzed the structure and stability of possible phases of a neutral electron-hole bilayer in a wide range
of the system's control parameters: layer separation, particle density,
and mass ratio at a fixed low temperature.
To this end we performed large-scale unbiased path integral Monte Carlo simulations. The properties of the observed exciton gas, exciton crystal, hole crystal and electron-hole plasma phases are characterized by pair distribution functions and snapshots of the instantaneous particle densities.
For a representative set of mass ratios, $M=1$, $10$, and $100$, the phase diagram was mapped out. It shows the complex interplay between the intra-layer and inter-layer correlations as well as stabilization and destabilization effects of the second layer on the Wigner crystal in a 2D Coulomb system. Currently, extensive simulations are underway that use a denser grid of simulation parameters that will allow to draw the phase boundaries with higher accuracy. 

Our results are important for strongly correlated two-component charged particle systems in general \cite{sccs_2011} and for complex plasmas in particular \cite{meichsner_cpp12,melzer_cpp12}.
Common to all these systems are collective phenomena resulting in a rich variety of phases. Of particular interest is the possibility to  control the interparticle interaction in e-h bilayers by changing external parameters (inter-layer separation, density). These concepts are also of interest for complex plasmas and for charged particles in contact with surfaces \cite{bronold_cpp12} where similar pinning and crystal stabilization effects occur due to the presence of surface defects \cite{bonitz_cpp12}.


\begin{acknowledgement}
  This work was funded by the Deutsche Forschungsgemeinschaft, SFB TR-24, projects A5 and A7.
  Numerical simulations were performed on the tera-flop PC-cluster at the Institute of Physics, University Greifswald.
  J.~Schleede acknowledges the hospitality of the Institute for Theoretical Physics and Astrophysics,  Christian-Albrechts-Universit\"at zu Kiel.
\end{acknowledgement}



\end{document}